# Bidirectional Quantum Controlled Teleportation by Using Five-qubit Entangled State as a Quantum Channel

Moein Sarvaghad-Moghaddam[1,*], Ahmed Farouk[2], Hussein Abulkasim[3]

*Abstract*— **In this paper, a novel protocol is proposed for implementing BQCT by using five-qubit entangled states as a quantum channel which in the same time, the communicated users can teleport each one-qubit state to each other under permission of controller. The proposed protocol depends on the Controlled-NOT operation, proper single-qubit unitary operations and single-qubit measurement in the Z-basis and X-basis. The results showed that the protocol is more efficient from the perspective such as lower shared qubits and, single qubit measurements compared to the previous work. Furthermore, the probability of obtaining Charlie's qubit by eavesdropper is reduced, and supervisor can control one of the users or every two users. Also, we present a new method for transmitting n and m-qubits entangled states between Alice and Bob using proposed protocol.**

*Index Terms*— **Bidirectional Quantum Teleportation, One-qubit State, Entanglement State, Five-qubit Channel.**

## I. Introduction

Quantum information theory represents a fundamental principle to understand the nature of quantum mechanics. The most used quantum principles are quantum teleportation and dense coding. In quantum teleportation (QT), the quantum information can be transmitted between remote parties based on both classical communication and maximally shared quantum entanglement among the distant parties [13, 14, 15, 16]. The main idea of QT was proposed when Bennett et al. [8] utilized the principle of Einstein-Podolsky-Rosen (EPR) pair as a quantum channel. After that, Zha et al. [9] demonstrated a new version of QT called as Bidirectional Controlled Quantum Teleportation (BCQT) by employing a five-qubit cluster state as a quantum channel. Subsequently, many development and research in BQCT area have been represented [10-16]. BCQT protocol allowed the communicated users, in the same time, to teleport the secret messages to each other under permission of a controller. A BQCT protocol Using the six-qubit state as a quantum channel has been introduced in [12,13] for transmitting one-qubit state in which the operation contained two-qubit measurements and so are not optimal. In [17], a BCQT protocol is proposed using the five-qubit state as a quantum channel for transmitting one qubit state between users to each other simultaneously. Unfortunately, the proposed protocol required additional quantum and classical resources so that the users required two-qubit measurements and applying global unitary operations. There are many approaches and prototypes for the exploitation of quantum principles to secure the communication between two parties and multi-parties [18, 19, 20, 21, 22]. While these approaches used different techniques for achieving a private communication among authorized users, but most of them depend on the generation of secret random keys [23, 2]. In this paper, a novel protocol is proposed for implementing BQCT by using five-qubit entanglement state as a quantum channel. In this protocol, users may transmit an unknown one-qubit quantum state to each other in the same time. The proposed BQCT protocol only use Controlled-NOT operation, proper single-qubit unitary operations and single-qubit measurement in the Z-basis and X-basis. In addition, controller as a supervisor can select the control of one of the users or every two users. Finally, this method is generalized for transmitting n and m-qubits entangled states between users.

The rest of the paper is organized as follows. Section II presents the proposed method in details. Preparation of five-qubit entangled state has been demonstrated in section III. The proposed work compared to previous work in section IV. Finally, Section V concludes the paper.

## II. Method

Consider a one-qubit state that Alice and Bob would like to teleport with each other, simultaneously, in the presence of Charlie as controller or supervisor is given by (1):

$|\Phi\rangle_A = \alpha_0|0\rangle + \alpha_1|1\rangle$ where $|\alpha_0|^2 + |\alpha_1|^2 = 1$

$|\Phi\rangle_B = \beta_0|0\rangle + \beta_1|1\rangle$ where $|\beta_0|^2 + |\beta_1|^2 = 1$

The protocol consists of the following steps:

**Step 1**: A quantum channel is shared among Alice, Bob, and Charlie with five qubits. According to the distribution of Charlie's qubit in the channel (c), the structure of shared channel can change as one of four states shown in Table I. As stated in this Table; Charlie can encode the four states with two

*Corresponding Author

Moein Sarvaghad-Moghaddam is with the Young Researchers and Elite Club, Mashhad Branch, Islamic Azad University, Mashhad, Iran (e-mail: moeinsarvaghad@mshdiau.ac.ir).

Ahmed Farouk is with Department of Physics and Computer Science, Wilfrid Laurier University, Waterloo, Canada (email: afarouk@wlu.ca ).

Hussein Abulkasim is with Faculty of Science, Assiut University, New Valley branch, Egypt Faculty of Science, South Valley University, Qena, Egypt (hussein@svu.edu.eg).

classic bits. Where the qubits $a_0, a_1$ and $b_0, b_1$ belong to Alice and Bob, respectively. Also, the qubit c regard to Charlie.

TABLE I
DIFFERENT CHANNELS CREATED USING OF DISTRIBUTION CHARLIE'S QUBIT.

| Encode | | Channel $|\psi\rangle_{(a_0)(b_0)(a_1)(c)(b_1)}$ |
|---|---|---|
| 0 | 0 | $\frac{1}{2}(|00000\rangle + |00101\rangle + |11000\rangle + |11101\rangle)$ |
| 0 | 1 | $\frac{1}{2}(|00000\rangle + |00111\rangle + |11000\rangle + |11111\rangle)$ |
| 1 | 0 | $\frac{1}{2}(|00000\rangle + |00101\rangle + |11010\rangle + |11111\rangle)$ |
| 1 | 1 | $\frac{1}{2}(|00000\rangle + |00111\rangle + |11010\rangle + |11101\rangle)$ |

For example, suppose that the channel shared among Alice, Bob and Charlie is encoded with 01 as shown in Eq. (2). Continue of the protocol is explained using the channel described in Eq. (2).

$$|\psi\rangle_{(a_0)(b_0)(a_1)(c)(b_1)} = \frac{1}{2}(|00000\rangle + |00111\rangle + |11000\rangle + |11111\rangle) \quad (2)$$

The general state of system is stated by Eq. (3).
$$|\phi\rangle_{(a_0)(b_0)(a_1)(c)(b_1)AB} = |\psi\rangle_{(a_0)(b_0)(a_1)(c)(b_1)} \otimes |\Phi\rangle_A \otimes |\Phi\rangle_B \quad (3)$$

**Step 2**: In this step, a Controlled-NOT gate is applied by Alice and Bob with A and B as control inputs and $a_0, b_1$ as target inputs, respectively. The general state of the system is stated by Eq. (4).

$$|\phi\rangle_{(a_0)(b_0)(a_1)(c)(b_1)AB} =$$
$$\frac{1}{2}[\alpha_0\beta_0(|00000\rangle + |00111\rangle + |11000\rangle + |11111\rangle)|00\rangle +$$
$$\alpha_0\beta_1(|00001\rangle + |00110\rangle + |11001\rangle + |11110\rangle)|01\rangle +$$
$$\alpha_1\beta_0(|10000\rangle + |10111\rangle + |01000\rangle + |01111\rangle)|10\rangle +$$
$$\alpha_1\beta_1(|10001\rangle + |10110\rangle + |01001\rangle + |01110\rangle)|11\rangle] \quad (4)$$

**Step 3**: Alice and Bob perform a single-qubit measurement in Z-basis on qubits $a_0, b_1$ respectively. The other qubits collapse to one of four possible states with the same probability (see Table II).

TABLE II
MEASUREMENT RESULTS BASED ON Z AND COLLAPSED STATES.

| Alice's Result | Bob's Result | The collapsed state of qubits $(b_0)(a_1)(c)AB$ |
|---|---|---|
| 0 | 0 | $\alpha_0\beta_0|00000\rangle + \alpha_0\beta_1|01101\rangle + \alpha_1\beta_0|10010\rangle + \alpha_1\beta_1|11111\rangle$ |
| 0 | 1 | $\alpha_0\beta_0|01100\rangle + \alpha_0\beta_1|00001\rangle + \alpha_1\beta_0|11110\rangle + \alpha_1\beta_1|10011\rangle$ |
| 1 | 0 | $\alpha_0\beta_0|10000\rangle + \alpha_0\beta_1|11101\rangle + \alpha_1\beta_0|00010\rangle + \alpha_1\beta_1|01111\rangle$ |
| 1 | 1 | $\alpha_0\beta_0|11100\rangle + \alpha_0\beta_1|10001\rangle + \alpha_1\beta_0|01110\rangle + \alpha_1\beta_1|00011\rangle$ |

**Step 4**: First, Alice and Bob notify the Z-basis measurement results to each other. Then, they apply X unitary operation to their unmeasured qubits $((b_0)(a_1))$ (see Table III). The state of the unmeasured qubits will be transferred to Eq. (5).

$$\alpha_0\beta_0|00000\rangle + \alpha_0\beta_1|01101\rangle + \alpha_1\beta_0|10010\rangle + \alpha_1\beta_1|11111\rangle \quad (5)$$

**Step 5**: Single-qubit measurements are applied in the X-basis on sending qubits $(A,B)$ by Alice and Bob. The other qubits collapsed to one of four possible states with the same probability (see Table IV).

**Step 6**: Alice and Bob notify their measurement results to each other. Then, they apply Z unitary operation to their unmeasured qubits $(b_0, a_1)$ as shown in Table V. The state of the unmeasured qubits will be in the state of Eq. (6).

$$(\alpha_0\beta_0|000\rangle + \alpha_0\beta_1|011\rangle + \alpha_1\beta_0|100\rangle + \alpha_1\beta_1|111\rangle)_{(b0)(a1)(c)} \quad (6)$$

TABLE III
APPLYING X UNITARY OPERATION

| Alice's Result | Bob's Result | Unitary Operation on $(b0)(a1)$ |
|---|---|---|
| 0 | 0 | $I \otimes I$ |
| 0 | 1 | $I \otimes X$ |
| 1 | 0 | $X \otimes I$ |
| 1 | 1 | $X \otimes X$ |

TABLE IV
THE MEASUREMENT RESULTS BASED ON X AND COLLAPSED STATES.

| Alice's Result | Bob's Result | The collapsed state of qubits $(b0)(a1)(c)$ |
|---|---|---|
| + | + | $\alpha_0\beta_0|000\rangle + \alpha_0\beta_1|011\rangle + \alpha_1\beta_0|100\rangle + \alpha_1\beta_1|111\rangle$ |
| + | − | $\alpha_0\beta_0|011\rangle - \alpha_0\beta_1|000\rangle + \alpha_1\beta_0|111\rangle - \alpha_1\beta_1|100\rangle$ |
| − | + | $\alpha_0\beta_0|100\rangle + \alpha_0\beta_1|111\rangle - \alpha_1\beta_0|000\rangle - \alpha_1\beta_1|011\rangle$ |
| − | − | $\alpha_0\beta_0|111\rangle - \alpha_0\beta_1|100\rangle - \alpha_1\beta_0|011\rangle + \alpha_1\beta_1|000\rangle$ |

TABLE V
APPLYING Z UNITARY OPERATION

| Alice's Result | Bob's Result | Unitary Operation on $(b_0)(a_1)$ |
|---|---|---|
| + | + | $I \otimes I$ |
| + | − | $I \otimes Z$ |
| − | + | $Z \otimes I$ |
| − | − | $Z \otimes Z$ |

**Step 7**: Charlie notify the distribution status of his qubit to Alice and Bob with two classical bits (see Table I). Then, he measures his qubit in X-basis and informs Alice and Bob of his result. If Charlie's measured result is $|+\rangle$ ($|-\rangle$), then, the states of other qubits are equal to either Eq. (7) or (8), respectively. The measurement results on Charlie's qubit are shown in Table VI for the different channels of Table I.

$$(\alpha_0\beta_0|00\rangle + \alpha_0\beta_1|01\rangle + \alpha_1\beta_0|10\rangle + \alpha\beta_1|11\rangle)_{(b0)(a1)} \quad (7)$$

$$(\alpha_0\beta_0|00\rangle - \alpha_0\beta_1|01\rangle + \alpha\beta_0|10\rangle - \alpha_1\beta_1|11\rangle)_{(b0)(a1)} \quad (8)$$

**Step 8:** According to Charlie's results, Alice and Bob apply Z unitary operation (see Table VI). Finally, Alice and Bob can reconstruct the transmitted states again and the BQCT is successfully completed (see Eq. (9) and (10)).

$$|\psi\rangle_A = b_0|0\rangle + b_1|1\rangle \quad (9)$$

$$|\psi\rangle_B = a_0|0\rangle + a_1|1\rangle \quad (10)$$

TABLE VI
APPLYING Z UNITARY OPERATION FOR THE DIFFERENT CHANNELS SHOWED IN TABLE I.

| Coding bits showing different channels | Charlie's Results | The collapsed state of qubits $(b0)(a1)$ | Unitary Operation on $(b_0)(a1)$ |
|---|---|---|---|
| 00 | $\|+\rangle$ | $\|00\rangle + \|01\rangle + \|10\rangle + \|11\rangle$ | $I \otimes I$ |
|    | $\|-\rangle$ | $\|00\rangle + \|01\rangle + \|10\rangle + \|11\rangle$ | $I \otimes I$ |
| 01 | $\|+\rangle$ | $\|00\rangle + \|01\rangle + \|10\rangle + \|11\rangle$ | $I \otimes I$ |
|    | $\|-\rangle$ | $\|00\rangle - \|01\rangle + \|10\rangle - \|11\rangle$ | $I \otimes Z$ |
| 10 | $\|+\rangle$ | $\|00\rangle + \|01\rangle + \|10\rangle + \|11\rangle$ | $I \otimes I$ |
|    | $\|-\rangle$ | $\|00\rangle + \|01\rangle - \|10\rangle - \|11\rangle$ | $Z \otimes I$ |
| 11 | $\|+\rangle$ | $\|00\rangle + \|01\rangle + \|10\rangle + \|11\rangle$ | $I \otimes I$ |
|    | $\|-\rangle$ | $\|00\rangle - \|01\rangle - \|10\rangle + \|11\rangle$ | $Z \otimes Z$ |

## A. Entangled state

For transmitting $n$ and $m$-qubits entangled states between Alice and Bob by utilizing the property of the proposed protocol, we will follow the same steps and continues to steps 9 and 10.

**Step 9**: Alice and Bob can notify the number of entangled states to each other by employing both the classical bits and coding table (see Table VII). While $\delta$ is a general changeable variable and $A$, $B$ correspond to Alice and Bob. The number of qubits for the entangled state can be determined by binary numbers sent by Alice and Bob to each other (see Table VII).

**Step 10**: According to the coding bits received by Alice and Bob, the circuit shown in Fig. 1 is applied to the received qubits ($|\psi\rangle_A$ and $|\psi\rangle_B$) by Alice and Bob. Subsequently, the entanglement states are considered successfully generated and the BQCT is effectively accomplished.

TABLE VII
ENCODING DIFFERENT ENTANGLEMENT STATES BY ALICE AND BOB.

| encode | The corresponding entanglement states |
|---|---|
| $0\ldots00$ | $\|\psi\rangle_\delta = \delta_0\|0\rangle + \delta_1\|1\rangle$ |
| $0\ldots01$ | $\|\psi\rangle_\delta = \delta_0\|00\rangle + \delta_1\|11\rangle$ |
| $0\ldots10$ | $\|\psi\rangle_\delta = \delta_0\|000\rangle + \delta_1\|111\rangle$ |
| $2^n - 1$ | $\|\psi\rangle_\delta = \delta_0\|0\ldots0\rangle + \delta_1\|1\ldots1\rangle$ |

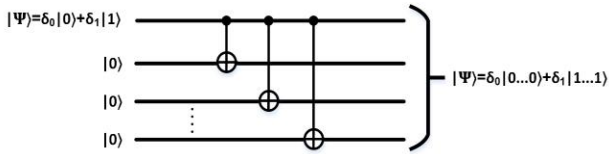

Fig. 1. Illustration of applying circuit by Alice and Bob for different entanglement states shown in Table VII.

## III. PREPARATION OF FIVE-QUBIT ENTANGLED STATE

Fig. II showed a quantum circuit of the proposed quantum channel with a five-qubit entangled state. The proposed quantum channel can be practically feasible using two Hadamard gates and series of two to four CNOT gates.
The steps of creating the channel are explained in details as the following;
First, zero states create an initial state (see Eq. (11)).

$$|\Psi\rangle_0 = |0\rangle_{a0} \otimes |0\rangle_{b0} \otimes |0\rangle_{a1} \otimes |0\rangle_c \otimes |0\rangle_{b1}. \quad (11)$$

Afterward, Hadamard gates are applied to qubits $b0$ and $a1$ as in Eq. (12):

$$|\Psi\rangle_1 = (|0\rangle_{a0} \otimes \frac{(|0\rangle+|1\rangle)}{\sqrt{2}}_{b0} \otimes \frac{(|0\rangle+|1\rangle)}{\sqrt{2}}_{a1} \otimes |0\rangle_c \otimes |0\rangle_{b1}) \quad (12)$$

After that, CNOT gates are employed with qubits b0 and a1 as control qubits and qubits a0 and b1 as target qubits. The whole state of system is as the following:

$$|\Psi\rangle_1 \quad (13)$$
$$= CNOT(b0,b1)\,(|0\rangle_{a0} \otimes \frac{(|0\rangle+|1\rangle)}{\sqrt{2}}_{b0} \otimes \frac{(|0\rangle+|1\rangle)}{\sqrt{2}}_{a1} \otimes |0\rangle_c \otimes |0\rangle_{b1})$$
$$= \frac{1}{\sqrt{8}}(|00000\rangle + |00101\rangle + |11000\rangle + |11101\rangle)_{a_0 b_0 a_1 c b_1}.$$

Finally, the controller can apply a CNOT gate with function $U$ so that he can consider each combination of qubits $a0$ and $b1$ as a control and qubit $c$ as a target. So, he can generate four different entangled states as shown in Table I. Furthermore, type of used combination can be encoded by two classical bit according to sequence $a0, b1$. For example, if he will create corresponding state in Eq. (2), then he need to apply one CNOT gate with qubit $b1$ as control and qubit $c$ as target. So, the proposed channel can be created as Eq. (14).

$$|\psi\rangle_{(a0)(b0)(a1)(c)(b1)} = \frac{1}{2}(|00000\rangle + |00111\rangle + |11000\rangle + |11111\rangle) \quad (14)$$

As a stated above, a five-qubit state can be prepared and used as a quantum channel.

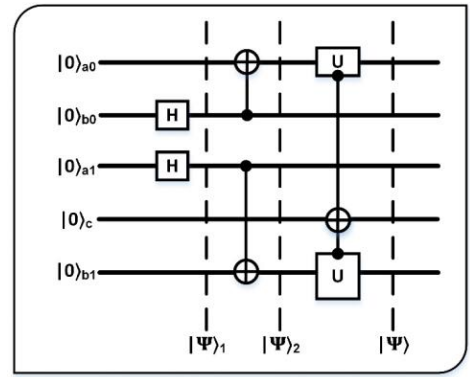

Fig. 2. Illustration of quantum circuit of proposed five-qubit channel.

## IV. COMPARISON

In this section, the proposed method is compared with some Previous works [12,13,17] in terms of the type of protocol, the number of qubits sent by Alice and Bob together, efficiency, BSMs (Bell-state measurements), SMs (single qubit measurements), and Prob. (i.e., the probability guess Charlie's qubit by an eavesdropper) as shown in Table VIII.

TABLE VIII
COMPARISON WITH PREVIOUS WORKS

| Methods | protocol | # Bob's qubits | # Alice's qubits | quantum channel | Efficiency | BSMs | SMs | Prob. |
|---|---|---|---|---|---|---|---|---|
| [12] | BCQT | 1 | 1 | Six-qubits | $\frac{1}{3}$ | 2 | 2 | $\frac{1}{4}$ |
| [13] | BCQT | 1 | 1 | Six-qubits | $\frac{1}{3}$ | 1 | 4 | $\frac{1}{2}$ |
| [17] | BCQT | 1 | 1 | Five-qubits | $\frac{2}{5}$ | 2 | 1 | $\frac{1}{2}$ |
| Our method | BCQT | 1 | 1 | Five-qubits | $\frac{2}{5}$ | 0 | 5 | $\frac{1}{4}$ |

In this table, the efficiency is defined [16] as the ratio of the number of sending qubits to the number of channel qubits. The proposed protocol has high efficiency compared to the previous works [12, 13]. While the proposed protocol and the protocol introduced in [17] have the same highest efficiency 2/5 but, Protocol [17] uses two global gates include of Quantum Controlled Phase Gate (QCPG) and controlled-not gate between Alice and Bob. For transmitting these gates, we require additional resources of entanglement states and qubits. Besides, our protocol reduces the probability of guess Charlie's qubit by an eavesdropper to 1/4 compared to [13, 17]. It is defined as the number of possible separate states obtained after measuring by Charlie and supervisor can control one of the users or every two users. The proposed protocol only use single-qubit measurement basis which is more efficient than two-qubit measurements (Bell state measurements) compared to previous works (see Table VIII). It is a well-known that Bell-state measurements can be decomposed into an ordering combination of a single-qubit Hadamard operation and a two-qubit CNOT operation as well as two single-qubit measurements.

## V. CONCLUSION

In this paper, a new protocol proposed for implementing BQCT by applying a five-qubit entanglement states as a quantum channel. Therefore, in the same time, the users can teleport arbitrary single-qubit state to each other with a supervisor Charlie. The implementation is mainly concerning the Controlled-NOT operation, proper single-qubit unitary operations and single-qubit measurement in the Z-basis and X-basis which provide the ability to realize it experimentally. The most important advantages of the protocol can be lowering shared qubits, single qubit measurements which are more efficient than two-qubit measurements and also, the probability of guess Charlie's qubit by eavesdropper is reduced, and supervisor can control one of the users or every two users. Finally, we presented a new method for transmitting n and m-qubits entanglement states between Alice and Bob using the proposed protocol.